\documentclass[%
 reprint,
 amsmath,amssymb,
 aps,
]{revtex4-1}

\usepackage{graphicx}% Include figure files
\usepackage{dcolumn}% Align table columns on decimal point
\usepackage{bm}% bold math
\usepackage{color}
\begin{document}

\preprint{APS/123-QED}

\title{Diffusion of self-propelled particles in complex media}% Force line breaks with \\

\author{Juan L. Aragones}
\affiliation{%
Department of Materials Science and Engineering, Massachusetts Institute of Technology, Cambridge, MA, 02139, USA.\\
Department of Theoretical Condensed Matter Physics, Universidad Autonoma de Madrid, Madrid, 28049, Spain.
}%
%\email{aragones@mit.edu}
\author{Shahrzad Yazdi}
\affiliation{%
Department of Materials Science and Engineering, Massachusetts Institute of Technology, Cambridge, MA, 02139, USA.
}%
\author{Alfredo Alexander-Katz}%
\affiliation{%
Department of Materials Science and Engineering, Massachusetts Institute of Technology, Cambridge, MA, 02139, USA.
}%
\email{aalexand@mit.edu}

\date{\today}

\begin{abstract}
The diffusion of active microscopic organisms in complex environments plays an important role in a wide range of biological phenomena from cell colony growth to single organism transport. Here, we investigate theoretically and computationally the diffusion of a self-propelled particle (the organism) embedded in a complex medium comprised of a collection of non-motile solid particles that mimic soil or other cells. Under such conditions we find that the rotational relaxation time of the swimming direction depends on the swimming velocity and is drastically reduced compared to a pure Newtonian fluid. This leads to a dramatic increase (of several orders of magnitude) in the  effective rotational diffusion coefficient of the self-propelled particles, which can lead to  "self-trapping" of the active particles in such complex media. An analytical model is put forward that quantitatively captures the computational results. Our work sheds light on the role that the environment plays in the behavior of active systems and can be generalized in a straightforward fashion to understand other synthetic and biological active systems in heterogenous environments.   
\end{abstract}

%\pacs{}% PACS, the Physics and Astronomy
%\keywords{active matter, squirmers}%Use showkeys class option if keyword

\maketitle

%\tableofcontents

%\s{}ection{Introduction}
The dynamics of an object in a fluid can be fully described using the time evolution of its speed and direction of motion. For small objects, such as microorganisms or micron-size particles, thermal fluctuations are also a critical ingredient by driving the translational and rotational diffusion. In particular, for living organisms or synthetic swimmers stochastic rotational diffusion determines the persistence of the resulting random walks such systems perform to sample the environment for a variety of functions, such as spreading diseases and nutrient uptake. For instance, the locomotion strategy of {\it E. Coli} bacteria follows a ``run-and-tumble" protocol to outrun diffusion~\cite{1977AmJPh..45....3P}. It can be described as a persistent random walk with ``run'' periods swimming in a straight line with nearly constant velocity, $U$, interrupted by sudden random change of their swimming direction with rate $\alpha$, known as ``tumble''. For other self-propelled microorganisms and synthetic active particles that swim without tumbling, the swimming direction changes gradually purely due to rotational diffusion of the particle characterized by the rotational diffusion parameter $D_{\theta}$. In dilute solutions of active particles in a Newtonian liquid, $D_{\theta}$ arises from rotational Brownian motion. For long time scales, one can show that the motion of self-propelled particles yields a random walk with effective translational diffusivity $D_t \propto U^2 / \alpha$ with $\alpha \leftrightarrow D_{\theta}$~\cite{solon2015active}.

Such scaling should break down when considering the dynamics of dense collections of active particles. In such crowded environments, many body interactions of cells with stochastic fluctuations produce interesting collective phenomena such as swarming~\cite{Bricard:2013jq}, pattern formation~\cite{Riedel:2005fz}, and motility-induced phase separation (MIPS)~\cite{Cates:2015ft}. In all these examples, diffusive properties can be modified by collisions and hydrodynamic interactions, that are highly influenced by the characteristics of surrounding media. Thus, we should also expect the diffusive properties of living organisms to be affected in their natural habitats, where microstructures such as polymers, particles, and passive entities determine the characteristics and mechanical properties of the background media ~\cite{2012PNAS..10913088H,Rabodzey:2008hl,Suarez:2005gc}. 

Even though the effect of complex environments on the locomotion mechanism~\cite{Patteson:2015et}, kinetics and collective behavior~\cite{Bozorgi:2011dw,Aragones:2016cna} of self-propelled particles has been widely studied, the effect on the stochastic properties, such as diffusion, in these systems has received far less attention. Recent experiments, however, have shown that the diffusion of an active Brownian particle~\cite{GomezSolano:2016ep} and {\it E. Coli}~\cite{Patteson:2015et} even in the dilute regime is highly influenced by the presence of polymers in the fluid solutions. The underlying mechanism of this behavior is yet to be understood. Here, we study theoretically and computationally the diffusion properties of self-propelled swimmers in complex media to shed light on how the environment affects the stochastic properties of the dynamics of such systems. In particular, we study a self-propelled particle embedded in a monolayer of passive (non-active) particles. 
We show that the diffusion properties of swimmers strongly depend on both the environment and the propulsion mechanism. 

% In the presence of a heterogeneous environment (e.g. a passive matrix), the rotational relaxation time of the swimming direction is significantly reduced with respect to the
% Brownian relaxation time. Additionally, the presence of passive particles enhances the effective drag on the swimmers, which results in a reduction of its swimming speed. Both factors lead to a drastic reduction of the translational diffusion of swimmers compared with the active Brownian case and should be considered when studying active particles in complex environments as could be dense solutions, viscoelastic media or interfaces. Finally, we put forward an analytical framework that captures quantitatively the numerical results and can be used to explain recent and future findings of the transport of active particles in complex environments. 
           
\begin{figure}[h!]
\centerline{\includegraphics[clip,scale=0.27,angle=-0]{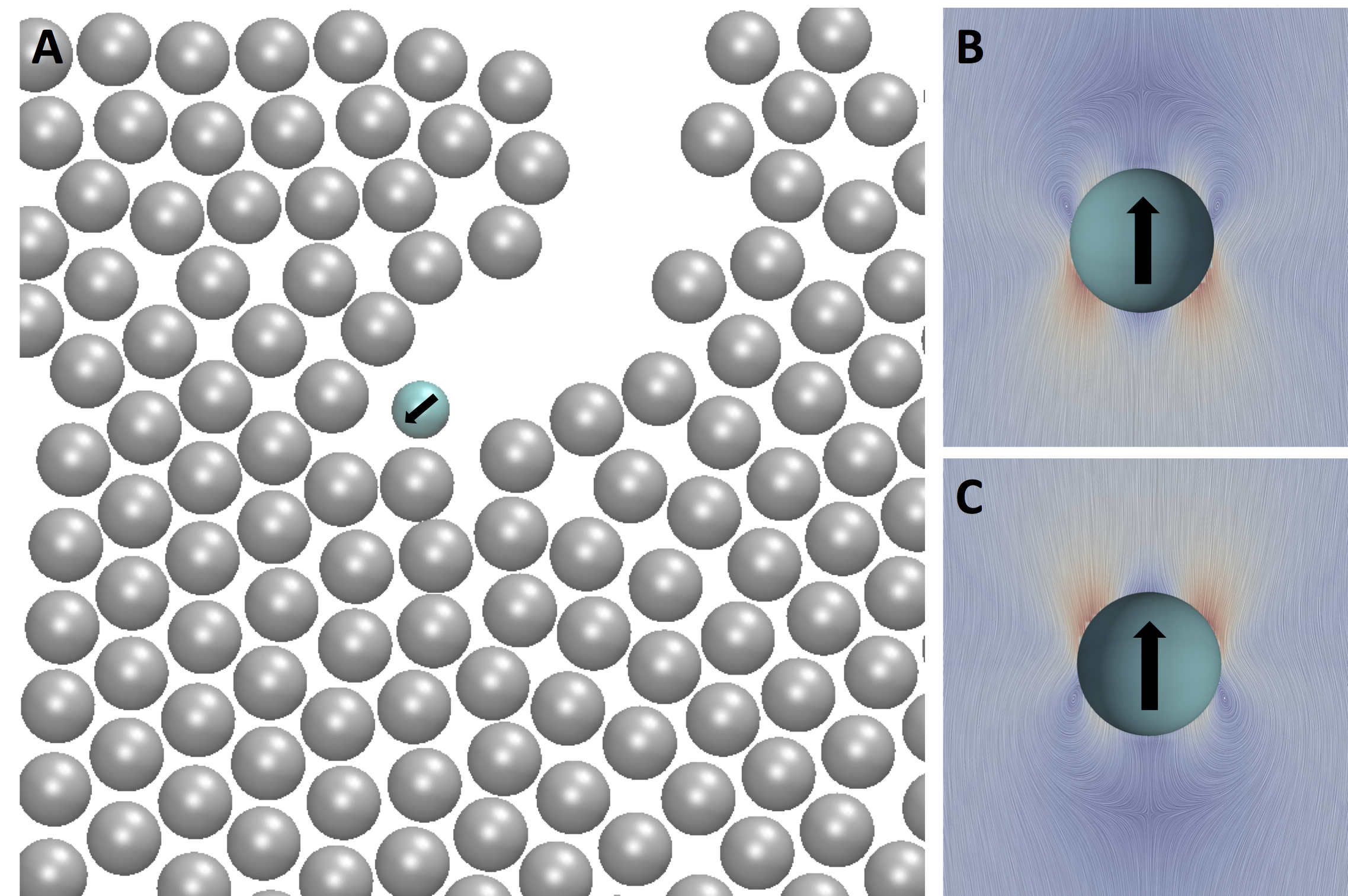}}
\caption{A) Schematic representation of a squirmer swimming within a passive monolayer. Velocity field generated by B) a pusher with $\beta$ = -5 and C) a puller with $\beta$ = 5.
\label{fig1}}
\end{figure}           
           
To investigate this system we perform numerical simulations of an active particle in a monolayer of passive particles using Lattice-Boltzmann (LB) method (see Fig. 1). As common practice, we set the fluid  density $\rho = 1$ and  viscosity $\eta = 1/6$. The simulation box is discretized in a three-dimensional grid with resolution $N_x \times N_y
\times N_z = 200 \times 200 \times 30$ bounded in the $z$-direction by no-slip walls and periodic boundary conditions in
$x$ and $y$-directions. The confinement in the $z$-direction mimics boundaries in biological 
environments and does not influence the conclusions from our work. We use 19-velocity model (D3Q19) and set the grid spacing $\Delta x$ and the LB time step $\Delta t$ equal to unity. The LB fluid is described by the fluctuating LB equation~\cite{Dunweg:2007km} with temperature $k_BT = 2 \times 10^{-4}$.
Interactions between the LB fluid and the particles are described by the bounce-back rule~\cite{Ladd:1994wb, Ding:2003wl}. In our
simulation model, all the particles (active and passive) have a diameter $\sigma = 10 \Delta x$ and do not overlap due to a short-range repulsive interaction of the form
$V(r_{ij}) = \epsilon \left( \frac{\sigma}{r_{ij}}\right)^{36}$, where $\epsilon = 1$ is the energy scale. The particles settle
on the bottom of the channel by applying a gravity force $F_G$ = 0.005. Under these conditions, a passive Brownian
particle has a translational diffusion coefficient given by
$D_t^{0} = \left( \frac{k_BT}{6\pi \eta a} \right) \left( 1 - \frac{9}{16} \frac{a}{h}\right) = 6.48 \times 10^{-6}\Delta x^2/\Delta t$,
where $a$ is the particle radius and $h$ the distance from particle surface to the wall~\cite{book:1310054}. Thermal fluctuations
also produce a rotational diffusion parallel to the wall described by
$D_{\theta}^{0}  = \left( \frac{k_BT}{8\pi \eta a^3} \right) = 3.82 \times 10^{-7}$ rad$^2$/$\Delta t$.

To model the active particles, we consider two different scenarios: one in which the active particle is propelled by a constant 
external force, $\mathbf{F}_A$. The unit swimming vector, $\mathbf{\hat{e}}$, of these externally driven (ED) particles is 
aligned with the applied force. For the second type of active particle, we use a prototyppical model, known as
``squirmer''~\cite{Blake:2004wl,Lighthill:1952ta}, in which the sphere is self-propelled via an effective tangential surface 
velocity on its surface, resembling the metachronal movement of cilia. In the truncated form, this velocity is expressed 
as~\cite{Spagnolie:2012gk,Zhu:2012ht}, 
$\mathbf{u}^s(\mathbf{\hat{r}}) = (\mathbf{\hat{e}} \cdot \mathbf{\hat{r}}\mathbf{\hat{r}} - \mathbf{\hat{e}})[B_1+B_2 \mathbf{\hat{e}} \cdot \mathbf{\hat{r}}]$,  
where $\mathbf{\hat{r}}$ is a unit position vector and $B_n(n = 1,2)$ is the $n$th mode of squirming velocity. The parameter
$\beta = \frac{B_2}{B_1}$ classifies the swimmers into pushers ($\beta < 0$) such as {\it E. Coli}, pullers ($\beta > 0$) such
as {\it Chlamydamonas}, and neutral ($\beta = 0$) such as {\it Paramecium}. 

We first study the motion of an active particle swimming in our confined system. At zero Reynolds number 
(Stokes regime), $Re = \frac{\rho U a}{\eta}$, the swimming speed of a squirmer in an unbounded Newtonian fluid is U $= 2B_1/3$.
Since the swimming direction in our simulation is limited to a plane parallel to the wall, the locomotion speed of force-free
squirmers is not affected by the wall. However, in the presence of inertia effects (Re $>$ 0.02), we observe an increase in
the speed of pushers, while pullers move slower (see Supplementary Information, SI). Our results are in agreement with
previous studies that used asymptotic analysis~\cite{KHAIR:2014bl} and numerically solved the Navier-Stokes 
equations~\cite{Chisholm:2016ja} (see SI). Unlike squirmers, the speed of ED particles strongly deviates from the Stokes
limit, due to higher drag force by the presence of the wall. Accounting for inertial and wall effects, our results are in close
agreement with theoretical predictions~\cite{book:1310054} (see SI). Using the mean square angular displacement (MSAD)
of the swimming direction, we observe that, in the absence of passive particles, the rotational diffusion coefficient for both 
squirmers and ED particles is independent of the swimming speed, U, and propulsion parameter, $\beta$, and is equal to
that of a passive Brownian particle, D$_{\theta}^{0}$ (see SI).

\begin{figure}[h!]
\centerline{\includegraphics[clip,scale=0.30,angle=-0]{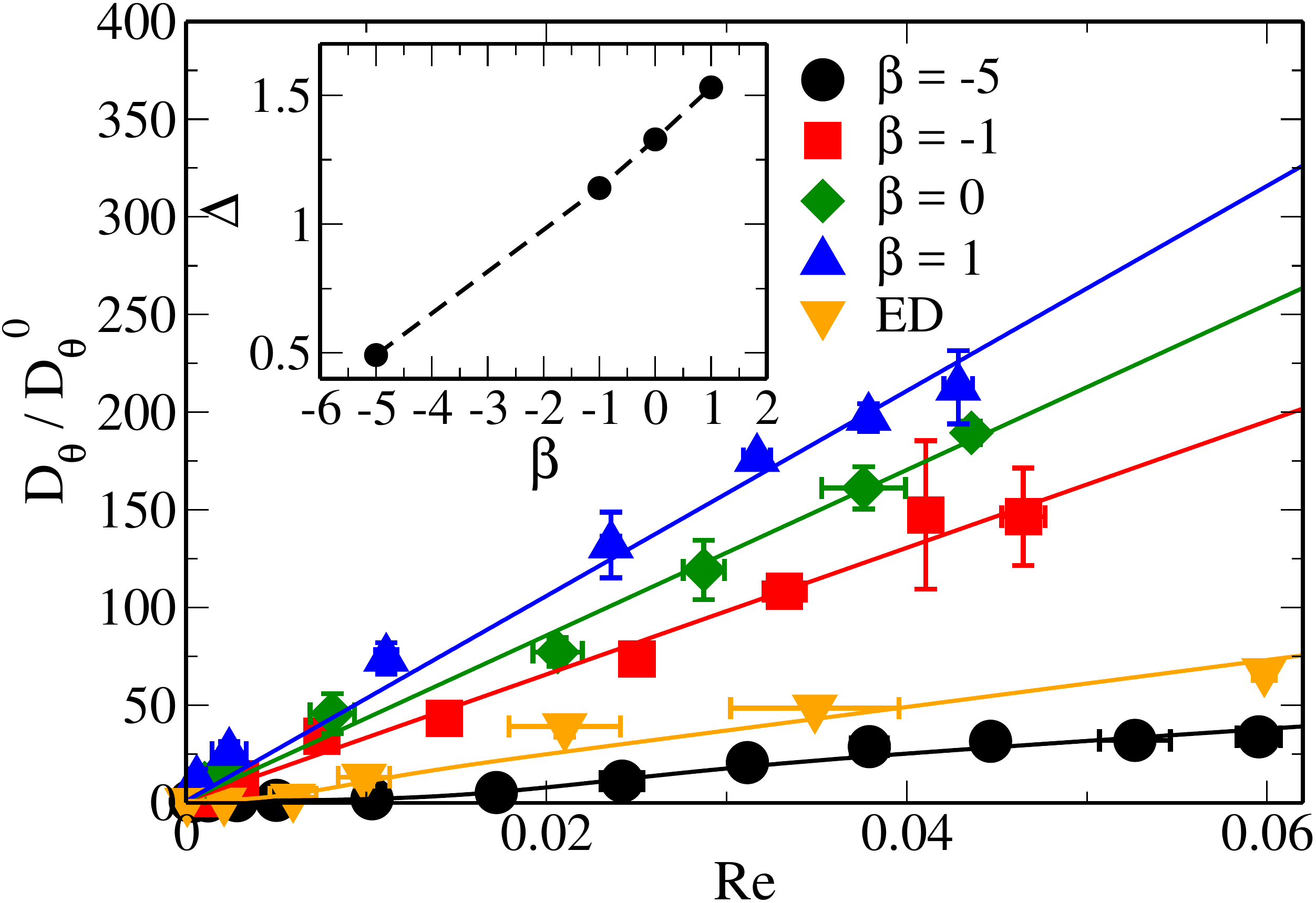}}
\caption{Rotational diffusion coefficient as a function of Reynolds number for different squirmer types and externally driven particles at $\phi_A$ = 0.5. The solid lines correspond to the analytical model described in the text. Inset: Phase shift, $\Delta$, as a function of $\beta$.}
\label{Dr_beta}
\end{figure}

In a Newtonian fluid and in the absence of other colloidal particles, the rotational diffusion of an active particle is due to thermal fluctuations of the fluid and thus, this type of active particles are usually referred as active Brownian (ab) particles.
Their translational diffusion at long time scales ($t >> 1/D_{\theta}$) is thus $D_t^{ab} = U^2/2 D_{\theta}$~\cite{Bechinger:2016cf}. On the other hand, active particles embedded in complex fluids such as viscoelastic media or colloidal suspensions can also experience athermal rotational relaxation. For example, in Fig.~\ref{Dr_beta} we show that the rotational diffusion coefficient of active particles embedded in a monolayer of passive particles at area fractions of $\phi_A$ = 0.5 
depends on the swimming velocity, Re, and propulsion parameter, $\beta$. As Re increases,
the rotational diffusion, D$_{\theta}$ for both squirmers and ED particles increases with respect to the Brownian rotational
diffusion, D$_{\theta}^0$. This enhancement is attributed to the increase in the stochastic torques on the active particle due to collisions with passive particles.

At a microscopic level, the rotation of the squirmer's swimming direction within a passive monolayer is controlled by the rate and effectiveness
of angular momentum  transfer via collisions. We consider that the athermal rotational
relaxation of the squirmer is driven by random collisions with the passive particles, which generate jumps, $\Delta$, in the
direction of movement, restricted to a plane parallel to the wall. Therefore, the orientational dynamics of the squirmers can
be described by the Debye rotational diffusion equation characterized by the exponential decay of the orientational correlation
function as $<P_l(\hat{e}(0) \hat{e}(t))> = \exp(-l^2 D_{\theta} t)$~\cite{Debye:tl}, where P$_l(x)$ is the Legendre polynomials and $l$ is a positive integer.
We assume that the occurrence of collisions between the squirmer and passive particles is described by a Poisson process
with a relaxation time $\tau$ (i.e. the characteristic jump interval), which is reversely correlated to the velocity of the squirmer, $U$, number area density of passive particles in the monolayer, $n = \frac{N}{A}$, and the effective length of collision (i.e. the
diameter of the squirmer), or cross section $\sigma_{AB}$; $\tau = 1/(U \sigma_{AB} n)$. Therefore, the orientational
correlation function in the long-time limit of this stochastic process is expressed by 
$<\hat{e}(0)\hat{e}(t)> = \exp[-\frac{t}{\tau}(1-<cos(\Delta)>)]$~\cite{Seki:2008kt,1986ApOpt..25.4585K}. 
This leads to an analytical expression for the rotational diffusion coefficient of the squirmer embedded in
monolayers of passive particles in terms of the jump magnitude and the collisional relaxation time, 
$D_{\theta}(Re) = D_{\theta}^0 + \frac{1}{\tau} \left( 1 - cos(\Delta) \right)$.
The solid lines in Fig.~\ref{Dr_beta} corresponds to the D$_{\theta}$(Re) obtained from this simple model, where the
phase shift, $\Delta$, is a free fitting parameter. This model captures the mean features of the orientational dynamics
of squirmers swimming in structured media, such as the increase of D$_{\theta}$ with the swimming velocity.
In addition, we observe that $\Delta$ increases with $\beta$, as 
shown in the inset of Fig.~\ref{Dr_beta}. This leads to a sharper increase of rotational diffusion for pullers. Not that in such a system, the rotational diffusion coefficient scales linearly with swimming velocity. For $\beta = 5$, D$_{\theta}/$D$_{\theta}^0$ increases up to 250 at $Re = 0.045$. Interestingly, we observe that for pushers with $\beta = -5$
the rate of collisions reaches a plateau below Re$ = 0.021$. A similar trend is observed for ED particles for Re $<0.006$.

\begin{figure}
\centerline{\includegraphics[clip,scale=0.325,angle=0]{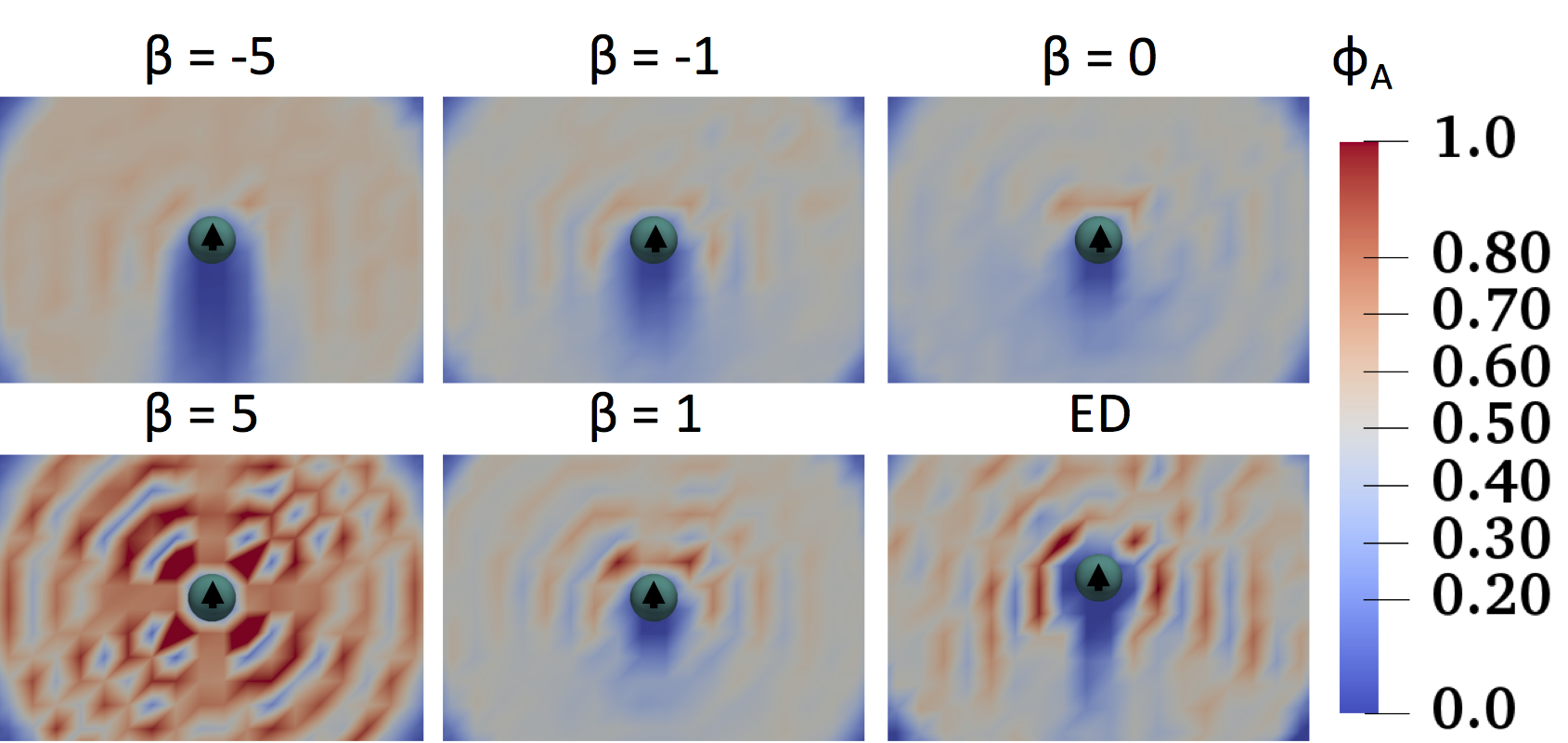}}
\caption{Color map of the passive particles area fraction, $\phi_A$, around the different type of active particles in $x-y$ plane swimming at about Re $= 0.12$: Pushers, pullers, neutral squirmers and ED particles. Blue regions corresponds to less dense areas, while red regions corresponds to more packed areas.
\label{rho_map}}
\end{figure} 

The velocity fields generated by each type of active particle determine the structure and dynamics of the surrounding passive particles, which control the rate and effectiveness of the collisions. ED particles translating in a viscous fluid experience drag forces, and the flow field that they generate can be represented by a Stokeslet (decays as $\sim 1/r$ ) in
far-field, whereas self-propelled particles do not experience net hydrodynamic drag. For a neutral squirmer, the generated
flow field is described as a source dipole (decays as $\sim 1/r^3$), while a stresslet (decays as $\sim 1/r^2$) can best describe
the flow field of pushers and pullers. In particular, pullers generate thrust by pulling the fluid from their front, while pushers
obtain momentum by pushing the rear fluid. Additionally, pullers or pushers with $|\beta|> 1$ produce recirculating flows (i.e.
axisymmetric vortices) in the rear or front of their body, respectively. The higher $|\beta|$, the greater the vorticity
~\cite{Chisholm:2016ja,Magar:2003wi}. The velocity fields generated by each of these propulsion mechanisms are shown 
in SI. We compute the average area density of passive particles around different active particles swimming at Re $\simeq 0.12$
in a monolayer with $\phi_A$ = 0.5, as shown in Fig.~\ref{rho_map}. The flow field generated by a puller with $\beta$ = 5
engages surrounding particles, which results in an increase of the area density of particles around the squirmer and consequently
impedes the motion of this type of swimmers in these complex environments (see Fig.~S5 and S8). The flow field exerted by
other squirmers with $\beta < 5$ generate regions depleted of passive particles on the rear part of their bodies, while
concentrating passive particles in their front. However, the vortices generated by pushers with $\beta$ = -5 repel passive
particles in their front, which results in a region of particles with reduced momenta and area density; this ultimately reduces
the rate of collisions. Similarly, the flow field around an ED particle reduces the density
of adjacent passive particles, and guides them to the sides of active particle. Due to this reduction of the rate and effectiveness
of the collisions, we observe that at small Re the D$_{\theta}$ of pushers with $\beta$ = -5 and ED particles is almost independent of the swimming velocity (see Fig~\ref{Dr_beta}). However, above a certain threshold, Re$_0$, 
where the active particle translates faster than the velocity at which passive particles in the front are repelled, collisions start
occurring and thus, the rate of
collisions increases with Re. To include this feature of the pushers with $\beta$= -5 and ED particles into our analytical model, 
we introduce a step function dependent on Re on the collision length, 
$\sigma_{AB}(Re) = \sigma_{AB} / (1+e^{-\kappa(Re-Re_0)})$, 
where $\kappa$ is the positive slope of Heaviside step function. The results of this model are shown in Fig.~\ref{Dr_beta} by
the black and orange solid lines.             

\begin{figure}[h!]
\centerline{\includegraphics[clip,scale=0.325,angle=-0]{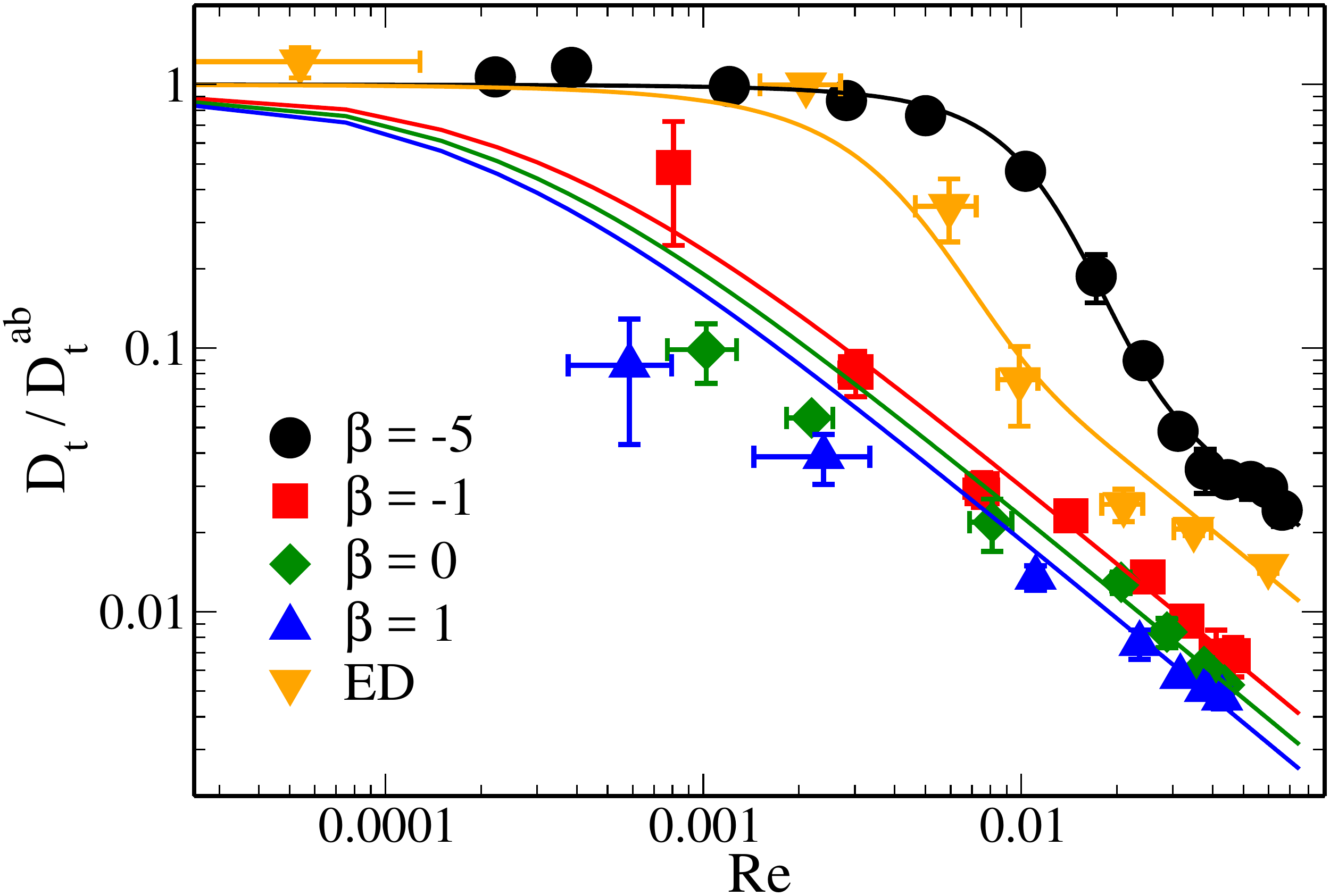}}
\caption{Translational diffusion coefficient as a function of Reynolds number for different squirmer types and externally driven particles. The squirmers are on the bottom wall of a channel of height $h$ = 30 $\Delta x$.
The D$_t$ are normalized by the translational diffusion coefficient of an active Brownian particle, 
$D_t^{ab} = U_0^2/(2D_{\theta}^0)$. The solid lines correspond to the analytical model described in the text.}
\label{Dt_beta}
\end{figure}

In addition to changing the orientational dynamics, the presence of a passive particles enhances the effective drag acting on
active particles, thereby reducing their swimming speed compared to that in Newtonian fluids (Fig.~S5). The presence of the 
dense media reduces the squirmer's velocity regardless of the swimming gait, except for pullers with $\beta$ = 5 for which
the speed reaches a plateau as Re increases. Reduction in both swimming speed and rotational relaxation time, results in a
significant decrease of the translational diffusion coefficient, D$_t$, with respect to the one of an active Brownian particle, 
D$_t^{ab}$, as shown in Fig.~\ref{Dt_beta}. The solid lines in Fig.~\ref{Dt_beta} corresponds to D$_t$ obtained from our
analytical model. These results show that the presence of a dense passive monolayer can induce the effective trapping
of squirmers and ED particles. Figure~S8 shows typical trajectories of different squirmer types within a passive monolayer
with $\phi_A$ = 0.5 over 10$^6$ time steps at Re $= 0.016$. We observed shorter trajectories as $\beta$ increases, which is due to the increase of rotational diffusion from pushers to pullers. This eventually leads to a significant decrease
of the translational diffusion of active particles within dense environments.
% We observe an increase in the curvature of trajectories with
% $\beta$, which implies an increase of the rotational diffusion from pushers to pullers.
%  Moreover, it shows a significant decrease
% of the translational diffusion of active particles within dense environments as $\beta$ increases.  

\begin{figure}[!h]
\centerline{\includegraphics[clip,scale=0.325,angle=-0]{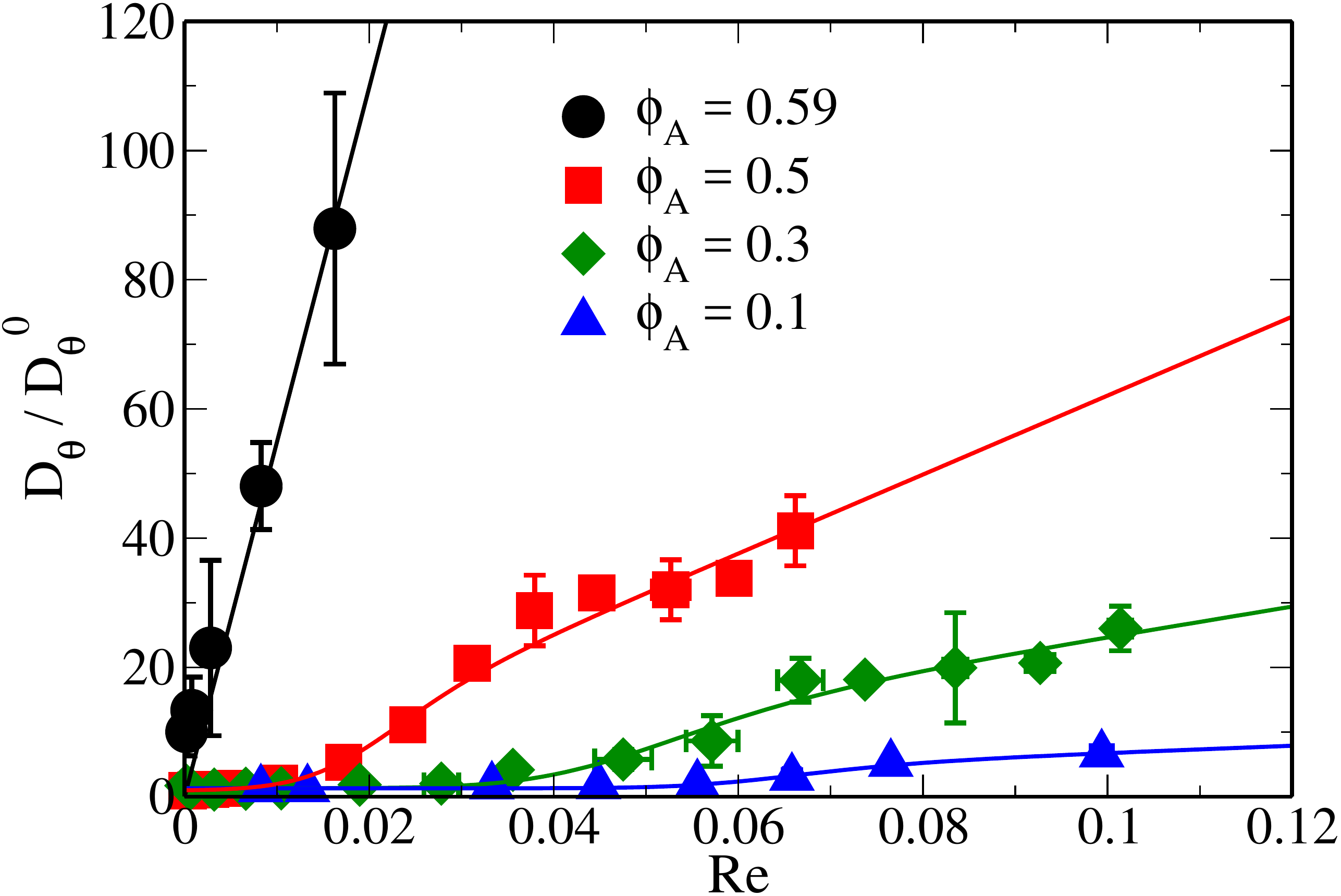}}
\caption{Normalized rotational diffusion coefficient as a function of the Re for
squirmers with $\beta$ = -5 when embedded in monolayers of passive particles at
different particle area fractions.}
\label{Dr_phi}
\end{figure}

Finally, we investigate the role of passive particle area fraction, $\phi_A$, on the swimming velocity and rotational diffusion (D$_{\theta}$) of a
pusher with $\beta$ = -5. The results of four different monolayer area fractions, $\phi_A$ = 0.59, 0.50, 0.30 and 0.1, show
that the swimming velocity of pusher reduces with $\phi_A$ (see Fig.~S6). Consistent with our model, we
observe that D$_{\theta}$ increases with the squirmer's swimming
velocity, as shown in Fig.~\ref{Dr_phi}. However, the increase of D$_{\theta}$ with Re strongly depends on $\phi_A$. At
small particle area fractions ($\phi_A <$ 0.59), the rate of increase for D$_{\theta}$ is slow when Re is below a critical value Re$_0$. However, above the critical Re$_0$, which reduces with $\phi_A$, rotational diffusion increases linearly with Re. At $\phi_A$ = 0.59, we only observe the linear regime of D$_{\theta}$
increasing with Re. The slope of D$_{\theta}$ with Re in the linear regime increases with $\phi_A$ and thus, we observe 
the highest effect for squirmers embedded in passive monolayers of $\phi_A$ = 0.59, where $D_r$ increases up to 90 times
compared to that of a Brownian rotational diffusion at the highest Re considered here.

In summary, we studied the role of environment and propulsion mechanism on the diffusion properties of swimmers. Squirmers swimming through a dense monolayer of passive particles experience an athermal 
rotational relaxation induced by the dissipative torques generated by particle collisions. These stochastic collisions result
in a random change of swimming direction that can be described by a Poisson process with a characteristic relaxation time
given by the squirmers' velocity and area fraction of passive particles, D$_{\theta}(Re,\phi_A)$. The rotational relaxation time
of the swimming direction decreases as the squirmer swimming speed and particle area fraction of the matrix increase. We also showed that the
effectiveness of angular momentum transfer during collisions, $\Delta(\beta)$, depends on the type of swimmer characterized by the parameter $\beta$. 
We observe that the orientational dynamic of the swimming direction of ED particles within these complex environments is similar to that of
pushers. The Stokeslet flow field generated by ED affects the structure and dynamics of the surrounding passive particles
similarly to the flow field of pushers with $\beta < -1$. Furthermore, the presence of the  passive matrix generates an effective
drag on the squirmers, which significantly reduces their swimming velocity with respect to the Newtonian fluid. Together with
the increase of the rotational diffusion coefficient, they lead to an effective trapping of squirmers within these complex 
environments. These results illustrate the complex dynamics of self-propelled particles within these passive environments,
which strongly depend on the characteristics of the media and propulsion mechanism. For example, we have seen that in the presence of more than one
squirmer within the passive monolayer, at a concentration of about 10\%, one can observe the formation of dense clusters of
passive particles. However, in contrast to previous numerical results~\cite{Stenhammar:2015ex}, the active particles do
not concentrate at the boundary of dense passive domains due to the increase in their rotational dynamics. These results show the importance of the propulsion mechanism of the active units on their dynamics in complex environments. Generalizations of this work may also  explain recent experimental observations of active colloids in polymer solutions \cite{GomezSolano:2016ep}. Moreover, our findings highlight the correlation between the propulsion mechanism of microswimmers and
the characteristics of the natural habitats, which may play a major role in determining their locomotion strategies.  
          
%\section { Acknowledgments }

This work was supported by the Department of Energy BES award \# ER46919.

%%\bibliography{scibib}
%%\bibliographystyle{./apsrev}
%\bibliography{./biblio}


\end{document}